\documentstyle[12pt,aaspp]{article}
\doublespace
\received{}
\accepted{}
\begin{document}

\def \GRBs{$\gamma$-ray bursts }
\def \GRBsb{$\gamma$-ray bursts}
\def \GRB{$\gamma$-ray burst }
\def \etal{{\it et al. }}
\vspace{0.6in}
\title{A Possible Explanation for The Peculiar Correlations in The Angular
Distribution of $\gamma$-ray Bursts}
\vspace{0.6in}
\author{Eyal Maoz}
\vspace{0.6in}
\affil{Harvard-Smithsonian Center for Astrophysics, \\
MS 51, 60 Garden Street, Cambridge, MA 02138}
\vspace{0.5in}
\centerline{E-mail: maoz@cfata4.harvard.edu}
\newpage
\begin{abstract}

It has been recently discovered that the angular autocorrelation function of
\GRBs exhibits sharp peaks at angular separations of
$\theta\!\lesssim\!4\deg$ (Quashnock and Lamb 1993), and at
$\theta\!\gtrsim\!176\deg$ (Narayan and Piran 1993).
These results were based on the publicly available BATSE catalogue.
While an excess of very close pairs of bursts can naturally arise from burst
repetition or from a spatial correlation of the burst sources,
a physical explanation for an angular correlation on a
scale of $\simeq\!180\deg$ seems inconceivable.

We show that the sharp peaks in the correlation function,
both at very small and at very large angular separations, can be
explained by a possible bias in the determination of the burst positions.
A generic way is described in which the suggested
bias can be introduced into the burst localization procedure,
either through instrumental imperfection or through the software analyses.

We apply Monte Carlo simulations to show that
the observed correlation function can be reproduced by the suggested effect.
We demonstrate that the results can nicely agree with the observations
even if only a fraction of the bursts are subject to the bias.

We emphasis that the only motivation for suggesting the existence of
this bias are the features
found in the angular autocorrelation function. It does not rule out
the possibility that bursts repeat. The natural way in which such
bias, if it exists, explains both sharp peaks, and the various conceivable
causes for the origin of this bias, make the bias hypothesis worth
considering.

\end{abstract}
%%%%%%%%%%%%%%%%%%%%%%%%%%%%%%%%%%%%%%%%%%%%%%%%%%%%%%%%%%%%%%%%%
\newpage

%\twocolumn

\section{Introduction}
In a recent analysis of the publicly available BATSE catalogue of \GRBs
Quashnock and Lamb (1993) have found a significant excess of
close pairs of bursts with angular separations $\lesssim 4\deg$. They
suggested that this is evidence for burst repetition on a
time scale of months,
where the scale of $4\deg$ reflects the localization error in the bursts'
directions, which is estimated to be of the same order (Fishman \etal 1993).
In a subsequent study of the burst angular autocorrelation function,
Narayan and Piran (1993) have discovered also a substantial correlation on
angular separations in the range
$176\deg\!\!<\!\theta\!\le \!180\deg$, comparable in
significance to the peak at $\theta\!\le\! 4\deg$ (Fig. 1). Lacking any
physical explanation for the excess of pairs separated by $\simeq\!180\deg$
they suggested that the two peaks in the correlation function, both at very
small and at very large angular separations, are either due to a statistical
fluctuation or caused by some unknown selection effect.

A correlation on scales $\lesssim\!4\deg$ can naturally arise
from burst repetition or from a spatial correlation of the burst sources.
In contrast, an astrophysical explanation for the angular correlation on
$\simeq\!180\deg$ scale is inconceivable. The sharpness of this peak (Fig. 1)
and the absence of any comparable excess of correlation in the entire
range of $4\deg\!\!<\!\theta\!<\!176\deg$ are especially puzzling.

In this paper we argue that the sharp peaks in the correlation function,
both on the $\lesssim\!4\deg$ and on the $\gtrsim\!176\deg$ scales, can be
explained by a possible bias in the determination of the burst positions.
In \S{2} we demonstrate how points can be distributed on a sphere ``almost
randomly'' while being strongly correlated at separations of $\sim\!0\deg$
and $\sim\!180\deg$, and on these scales only.
In \S{3} we describe a generic way in which the suggested
bias can be introduced into the burst localization procedure,
either through instrumental imperfection or through the software analyses.
In \S{4} we apply Monte Carlo simulations to show that
the observed correlation function can be reproduced by the suggested effect,
even if the bias is applied only to a fraction of the burst population.

\section{Angular Distributions On Rings}
First, let us discuss a purely mathematical problem which will be related to
the peculiar correlations in the burst distribution in \S{3}.

The angular two-point correlation function, $w(\theta)$, averaged over many
realizations of randomly distributed points on the celestial sphere, is
identically zero. It is easy to show that if, instead of an isotropic
distribution, the points are randomly distributed only within an angular
distance  $\Delta\theta/2$ from a single great circle on the sky (hereafter,
a ring of width $\Delta\theta$) then $w(\theta)$ will exhibit
pronounced peaks at
$\theta\!\le\!\Delta\theta$ and at $\theta\!\ge\!180\!-\!\Delta\theta$.
In fact, $w(\theta)$ would present such peaks even if the points were
distributed within a set of an arbitrary number of randomly oriented rings.
In order to demonstrate this we have constructed many Monte Carlo sets
of points with such underlying distributions, using the following procedure:
a) choose $k$ randomly oriented rings, each $4\deg$
in width. b) generate random positions on the sky and keep only those points
which fall within at least one of the rings.
We do it in such a way that the probability of selecting
a point in a certain direction is proportional to the
number of rings within which that direction is located, namely, there are
higher probabilities to directions where rings overlap.
For illustration, figure 2 presents one realization of randomly
distributed points within 3 rings, and within 60 rings.

For any given number of rings, $k$, we have constructed many realizations of
$260$ points using the above scheme (there are 260 bursts in the available
BATSE catalogue), and calculated $w(\theta)$ for each realization.
For any value of $k$ we find that $w(\theta)$, when averaged over
the ensemble of realizations,
sharply peaks at angular separations of $\lesssim\!4\deg$
and $\gtrsim\!176\deg$. Figure 3 shows the results for $k\!=\!20$, and
$k\!=\!60$.
The effect remains, though with a lower significance level, even when $k$
is so large that the rings practically cover the entire sky.
The reason is that any
number of randomly oriented rings are extremely unlikely to
cover the sphere {\it uniformly}, and the regions where rings
overlap are typically $\sim\!4\deg$ in size, which shows up in the
correlation function.

It is not difficult to trace the origin of the two peaks:
The correlation at $\theta\!\lesssim\!4\deg$ is due to the small width
of the rings and due to the fact that the enhanced probability regions (where
rings overlap) are of the same angular size. The peak at $\simeq\!180\deg$
is due to the ring-like structure of the underlying distribution, and due to
the fact that if two rings cross at a certain point then they must
cross also at the diametrically-opposed direction.
Thus, the enhanced probability regions themselves (which are typically
$\sim\!4\deg$ in size) are correlated on angular separation of exactly
$180\deg$.

Obviously, the underlying distribution function of any set of rings is
symmetric in opposite directions. If rings cross at an
angular distance $\phi$ from a certain position on the sphere, then their
second crossing point is at a distance of $180\deg\!-\!\phi$
from that position. Therefore, the
autocorrelation function of distributions on rings must be symmetric, i.e.,
satisfy $w(\theta)\!=\!w(180\!-\!\theta)$, a property which is indeed evident
in Figure 3.

As we shall see in \S{4}, the effect is weakened when such distribution is
considerably diluted with a truly isotropic distribution, but it
does not disappear altogether.

\section{The ``Ring Bias''}
The peculiar correlations in the burst distribution (Fig. 1) can be
understood if the bursts, or at least a substantial fraction of them,
are somehow biased to lie
within a set of narrow rings.  We shall now argue that the structure of
the BATSE instrument, the configuration of the spacecraft's orientations,
and the burst localization software
allow considerable room for a systematic localization error which tends to
slightly displace bursts, making their positions coincide with a certain
configuration of rings.
First, we mention the relevant basic characteristics of
the instrument and then describe a simple generic way in
which such a bias can be introduced, keeping in mind that the real cause for
such bias, if it exists, is likely to be more complex.

The BATSE experiment (Fishman \etal 1989) consists of eight Large Area
Detectors placed at the corners of the Gamma Ray Observatory spacecraft, with
an octahedron geometry.  The detectors are uncollimated and the location of
a burst is determined by comparing the relative count rates from the
separate detectors. A burst can be viewed by up to four detectors, but
bursts are triggered on-board when the count rate in the $50\hbox{-}300$
Kev energy range increases above the background
by $5.5\sigma$ or more in {\it at least two\/} detectors simultaneously.

The response of BATSE's detectors to incident flux is anisotropic, so the
sensitivity threshold for detecting a burst varies across the sky.
A weak burst at one location on the sky may create acceptable rates in
BATSE's detectors while an identical burst at another location will not
(Brock \etal 1991).
This known intensity-dependent effect is in GRO coordinates, but because
of the many orientations of GRO for the various observation periods,
this effect is believed to be averaged away when calculating dipole or
quadrupole moments of the burst distribution. However,
it may not average away when calculating the correlation function.

For example,
many bursts are likely to produce the required $\ge\!5.5\sigma$ signal only
in two detectors, either because the direction of a burst is close
to the plane defined by the two detectors (so the other detectors are at
higher angles of incidence), or simply because the burst is weak (most bursts
are weak ones). The important point is that if no other detector had fired
(say, due to its high angle of incidence),
or alternatively, if the burst localization software had ignored (or
mistreated)
a $<\!\!5.5\sigma$ signal from the other detectors, then the burst would be
determined to lie in the plane defined by the two detectors.
A burst's direction would ``collapse'' into a ring, rather then into the exact
plane, if, for example, the reason for the bias is that a low-amplitude signal
in a third detector was not taken into account properly. The width
of the ring would be $\simeq\!4\deg$, which is comparable to the already known
localization errors due to
statistical errors and imperfect modeling of the instrument.

Revealing the exact origin of such possible
localization error requires a high level
of technical understanding of the instrument's operation, and familiarity with
the relevant software. The following are just a few conceivable causes which
might be related to the suggested effect:
a) an insufficient response of the detectors
at incidence angles which approach $90\deg$. b) differences in the thresholds
of the eight detectors. c) the burst
localization software ignores detectors which fire with low significance
levels (or takes them into account incorrectly).
d) some correlation between the
spacecraft's different orientations (see \S{4}).

The suggested localization error slightly displaces bursts from their true
positions and puts them within one or a few rings which have fixed orientations
relative to the spacecraft. However, since GRO has changed its orientation
$\sim\!30$ times during the first year of observation, the relevant number of
rings is larger by a factor of $\sim\!30$ (but see \S{4}).
Furthermore, the rings should be unequally weighted since the GRO spent
different periods of time in its various orientations. Therefore,
we shall now derive the configuration of rings,
based on the GRO orientations in the first year of observation, and check
whether the discovered peaks in $w(\theta)$ can be reproduced by the suggested
bias.

\section{GRO's Rings}
During the time in which the first 260 bursts were detected
(from April 21, 1991 until March 5, 1992)
the GRO has changed its orientation 31
times, spending between 3 to 14 days in each orientation (the GRO's
pointing plan is publicly available).
Suppose that there is only one plane in
GRO coordinates towards which the positions of the bursts are shifted, and
make, at the moment, the (too optimistic) assumption that the bias applies
to all bursts. Instead of performing Monte Carlo simulations of 31 randomly
oriented rings, we ``attach'' a plane to the spacecraft and derive the
configuration of rings which corresponds to the set of orientations that
this plane had portrayed
 during the year of observation (hereafter, the GRO rings).
The orientation of the plane relative to the spacecraft is irrelevant as long
as it is fixed during the entire year. Now we generate simulated sets of 260
points by drawing randomly from this specific set of rings as described
in \S{2}, but with a probability of drawing a point within each ring
proportional to the time the spacecraft
spent in the corresponding orientation. Obviously,
the probability to draw a point in a certain direction is weighted by
the total period of time that this direction had been located within a ring.
The autocorrelation function for the GRO set of rings, averaged over many
realizations, is displayed in figure 4.

We find that the sharp peaks at $\lesssim\!4\deg$ and at $\simeq\!180\deg$
are even higher than the expected from a randomly oriented set of 31 rings,
and that there is also a marginally significant correlation on
intermediate angular
scales. We have examined the GRO pointing plan for the relevant
period of time and found that the GRO orientations were not distributed
at random. The spacecraft is usually oriented in ways that the other
on-board (collimated) experiments, e.g., OSSE and COMPTEL, are pointed towards
especially desirable targets like the Crab pulsar, the Galactic center,
CYG X-1, etc. Furthermore, the pointing plan is optimized, for example, in a
way that will enables the OSSE detector (which can move in the x-y plane of
the spacecraft) to alternate between two targets when one of them is
below the horizon (Eric Chipman, GRO Science Support Center; private
communication).  Chosing an orientation for the spacecraft is also
constrained by the requirement that the collimated detectors will not point
too close to the sun. Consequently, 6 of the GRO orientations are
redundant, and the others are correlated in some non-trivial way.
This, and the variance in the durations that the spacecraft spent in the
different orientations imply that the number of rings is effectively lower
than 25.

We have assumed so far that the ``ring bias'' applies to all bursts, but it
is reasonable to expect that only some fraction of them would be subject
to the effect. Lacking the exact mechanism which produces this bias (but
see \S{3}) we shall now show an example which demonstrates
that even if a biased distribution is
diluted with randomly (and isotropically) distributed bursts the observed
features in $w(\theta)$ can still be reproduced.
For that we have generated many simulated samples of 260 points, where the
position of each point was determined in the following way: a) draw a random
location from an isotropic distribution and choose a random day within the
year of observation. b) find the orientation of the plane (which is fixed to
the spacecraft) at that time, and calculate the angular distance of the point
from the plane. c) If the angular distance is less than $30\deg$ then the
position of the burst ``collapses'' to the corresponding ring. Otherwise,
the burst remains in its original position.  Figure 5 presents the
resulting $w(\theta)$, averaged over many such realizations.
It is consistent with the observed autocorrelation function (Fig. 1).

We point out that the quadrupole and dipole moments for such distributions
are {\it not\/} in conflict with the values obtained for the burst
distribution.

\section{Discussion}
The main conclusion of this paper is summarized in the abstract so we
avoid redundancy here. We shall just stress that the suggested bias,
if it exists,
does not rule out the possibility that burst sources repeat on a time
scale longer than a year, and it does not necessarily imply that the dipole and
quadrupole moments of the burst distribution are much different than their
current values.

I wish to thank Ramesh Narayan, Bill Press, George Rybicki, Tsvi Piran,
and John Dubinski for discussions and comments.
This work was supported by the U.S. National Science Foundation, grant
PHY-91-06678.

\newpage

{\bf Fig. 1} -  The angular autocorrelation function of the bursts in the
publicly available
BATSE catalogue, binned into intervals of $4\deg$ (Narayan and Piran 1993).
The $1\hbox{-}\sigma$ error bars were calculated using simulations of random
isotropic distributions (the data for these plots
 have been kindly provided by Ramesh Narayan and Tsvi Piran).
Notice the sharp peaks at $\theta\!\lesssim\!4\deg$ and at $\theta\!\gtrsim\!
176\deg$. $\:\:$ a) The full sample of 260 bursts. $\:$ b) Only those
131 bursts which have formal position errors smaller then $4\deg$.

{\bf Fig. 2} - a) An equal area projection of one realization of 260 points
which are distributed
randomly within 3 randomly oriented rings, according to the scheme
described in \S{2}. b) A similarly generated distribution but within 60 rings.
Although
a visual assessment cannot distinguish between this and a random isotropic
distribution, the angular autocorrelation function for these distributions
sharply peaks at very small and at very large angular separations
(see Fig. 3).

{\bf Fig. 3} - The angular autocorrelation function, averaged over many
realizations, for 260 points randomly distributed
within 20 randomly oriented rings (a), and within 60 rings (b).
The error bars reflect one standard deviation obtained from the ensemble of
realizations. Notice the sharp peaks at $\theta\!\lesssim\!4\deg$ and
at $\theta\!\gtrsim\!176\deg$, and the symmetry property $w(\theta)\!=\!
w(180-\theta)$ (see text). These functions appear smoother than those in figure
1 because they are the product of averaging $w(\theta)$ over many realizations.

{\bf Fig. 4} - The angular autocorrelation function for 260 points randomly
distributed on the GRO set of rings (see text). The peaks in the first and
last bins are even higher than the expected from a randomly oriented set
of 31 rings, which indicates that the effective number of rings is smaller.
Indeed, the various orientations of the spacecraft are not randomly
distributed (\S{4}), which is also reflected by the marginally
significant variations of $w(\theta)$ on intermediate angular scales.

{\bf Fig. 5} - Similar to figure 4, but now the distribution is considerably
diluted by a truly isotropic distribution (unbiased positions) (see text).

\end{document}